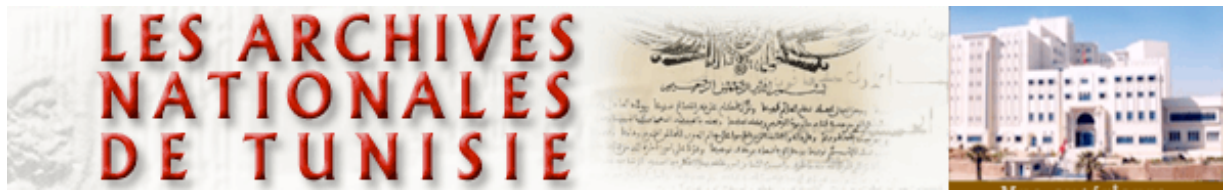

# Les valeurs linguistiques et culturelles des documents numériques et leur traitement sémantique dans les systè-mes d'information électroniques

**Mokhtar BEN HENDA**
ISD, Université La Manouba, TUNISIE
Mokhtar.Benhenda@isd.rnu.tn

*Résumé :*

Le document numérique évolue de façon rapide et spectaculaire dans sa structure et son contenu d'information véhiculés sur les réseaux et les systèmes d'information. Gé-néralement appréhendé comme support neutre d'une information porteuse d'une va-leur sémantique, le document numérique est pourtant porteur de paramètres qui déno-tent de certaines valeurs culturelles et linguistiques propres à son créateur.

Ce document essaie de mettre l'accent sur quelques aspects techniques qui traduisent cette identité intrinsèque des documents électroniques. L'objectif en est de définir un ensemble de paramètres et de recommandations capables de préserver au document numérique une identité culturelle et linguistique qui saura l'identifier auprès de ses utili-sateurs potentiels dans les systèmes d'information ouverts et distribués. Le tout vise en fin de comte à proposer des recommandations de « bonnes pratiques » pour les pro-ducteurs de documents numériques totalement ou partiellement arabe.

## 1. Introduction

La conscience collective nationale en Tunisie, comme partout ailleurs dans les pays en voie de développement et peut être aussi, même a des degrés moindres, dans les pays développés, est encore ancrée dans le référentiel papier. Affirmation trop hâtive peut être, mais qui certainement confirme que la théorie de la troisième vague d'Alvin Toffler et les prévisions de Tom Surprenant pour une société sans papier restent encore de l'ordre des chimères. Le document traditionnel puise encore une notoriété évidente de son support et de sa légitimité largement inspirée des sens, de la cognition et des usages : le toucher, la matérialisation et la portabilité. Le passage à l'ère numérique allait pourtant perturber cet ordre établi et pousser lentement vers une refonte des rapports dialectiques et fonctionnelles entre tous les acteurs de la société du savoir tant au niveau de la production des document, qu'au niveau de leur gestion, usage et conservation. Inscrits dans la durée, beaucoup de critères sont en train de métamorphoser l'univers de l'édition numérique et des techniques et procédures qui lui sont associées.

L'une des phases initiales de la numérisation des documents est sans doute l'étape de la codification des caractères. Du temps des données exclusivement textuelles, il était nécessaire de s'approprier cette technique pour entrer dans l'univers des documents électroniques produits et gérés par une nouvelle technologie numérique naissante. Une fois rodée et établie, la numérisation s'est ensuite préoccupée du rétroactif analogique dans une optique de numérisation massive des fonds des documents papiers. Les projets de numérisation et de reconnaissance optique des caractères des collections documentaires, des fonds d'archives et des séries de périodiques en format papier ont fait l'unanimité à échelles nationales et internationales. Avec l'accroissement de la masse documentaire papier et électronique, il était inévitable de trouver des alternatives techniques pour mieux contrôler cette masse de documents à des fins de capitalisation et de mutualisation. Le principe des métadonnées est ainsi réapparu à travers le réseau Internet pour constituer l'un des fondateurs clés des nouveaux systèmes d'information électroniques et des services à valeurs ajoutées qui en découlent comme le workflow et le datamining. Cette nouvelle approche de gestion des documents électroniques s'est également enrichie des réalisations parallèles dans le domaine de l'édition électronique qui s'est vu renforcé par les apports des techniques des langages structurés comme SGML et XML. Cette évolution est à la base de ce que l'on identifie aujourd'hui de réseaux sémantiques construits autours de la technologie XML pour ce qui est des structures, des ontologies pour ce qui des contenus et des parseurs intelligents pour la reconstitution des textes et leurs adaptations à des accès personnalisés.

Le développement des systèmes d'information et des outils associés a fait ainsi apparaître plusieurs problématiques du genre relations entre la mise en forme des documents et leur perception au niveau des interfaces homme machine (problèmes ergonomiques), la construction des documents dynamiques, l'archivage et ses rapports dynamiques avec la publication, le suivi des processus de validation, d'authentification et de génération automatique des documents, les liens et les facteurs d'impacts entre les documents… Bref, toute la panoplie de rapports entre la production, la gestion, le partage et l'exploitation des ressources électroniques. En d'autres termes, toute la logique du métier de production et de savoir faire de médiation se trouve affectée aujourd'hui par une nouvelle acception du documents dans ses dimensions physiques, techniques, scientifiques, juridiques et commerciales. Les documents structurés sont aujourd'hui à la mode des systèmes automatisés d'information. Ils s'associent en grande partie aux principes des bases de données dynamiques, centralisées ou décentralisées, qui constituent la source première de l'ensemble des fichiers des documents électroniques malgré leurs configurations variées. « *Un document n'aurait alors d'existence à proprement parler qu'à deux moments : éventuellement celui de sa conception par son auteur qui devra le visualiser ou l'entendre, pour s'assurer qu'il correspond à ses choix et surtout celui de sa reconstruction par un lecteur* »[ **Salaün, 2004**].

Dans cette introduction, qui n'a d'objectif que de cerner le contexte global des documents électroniques et des nouvelles conditions de leurs genèses, gestion et usages, notre intérêt portera sur un aspect d'une dimension particu-lièrement afférente aux contenus des documents électroniques : leurs valeurs linguistiques et culturelles comme vecteurs déterminants dans les stratégies de leur conception, usage et conservation à long terme dans le contexte socioculturel et technologique propre à une communauté quelconque. Notre intérêt sera porté essentiellement sur le contexte tunisien et la particularisme des documents numériques en langue arabe : spécificités, évolution et perspectives.

## 2. Le contexte technologique tunisien :

Tous les indicateurs démontrent que la Tunisie est désormais entrée de plein pied dans l'ère de la société de l'information à travers les processus et les projets en cours qui véhiculent des valeurs de transition rapide d'un stade historique de développement agraire, furtivement doublée par une mise à niveau industrielle de base, vers une refonte sociétale basée sur des acteurs de développement économique bâtis autour de la prestation de services et de l'économie de l'immatériel. En moins d'une décennie, le contexte tunisien a transcendé toutes les prévisions du début des années 90 en termes de services d'information basés sur un usage poussé des technologies de l'information et de la communication (TIC). La médiation électronique, sur site ou en ligne, est devenue une composante essentielle de la réalité sociale tunisienne. Les TIC connaissent une extension de plus en plus large et de plus en plus cohérente, renforcées par des programmes de socialisation multiples, comme « l'ordinateur familial », les abonnements Internet au tarif famille, les incubateurs technologiques, les cybercafés etc. Ce qui engendre des coûts toujours plus bas à des performances égales, une interopérabilité et une convergence multimédia de plus en plus globale, une logique de l'intégration de services par les réseaux. Toutes proportions gardées d'une contextualisation géopolitique et économique mondiale, la Tunisie figure parmi les pays en voie de développement les plus nantis en moyens TIC et en ressources et services d'information scientifique et technique (IST). Avec ses dix millions d'habitants, la Tunisie offre une densité téléphonique de 25 lignes pour 100 habitants (en 2004), un réseau téléphonique entièrement numérisé (6500 km en fibres optiques) depuis 1999, 55 boucles SDH permettant un débit de 10 à 40 Gbits/s, deux opérateurs de téléphones mobiles d'une capacité de 1500.000 lignes, deux passerelles internationales, des liaisons sous marines avec l'Europe et des liaisons spatiales numériques sur Internet et via les satellites Arabsat, Orascom, Thuraya. Douze fournisseurs de Services Internet disposent d'une bande passante à l'international d'une capacité de 155 Mbits/s. Les télécommunications ont inexorablement été l'acteur crucial pour amorcer un élan stratégique considérable dans le développement des services d'information et de communication en Tunisie. Tous les secteurs névralgiques de la société ont été ainsi touchés, à des écarts variables, par une empreinte de connectivité technologique qui leur a permis d'œuvrer pour des perspectives de partage, d'efficacité et de rentabilité meilleures. Des systèmes d'information comme INSAF (application des personnels de l'Etat), ADEB (application de gestion des dépenses au Budget de l'Etat), SINDA (application de gestion douanière), La Liasse Unique Tunis Trade Net (échanges avec l'extérieur), MADANIA (réseau national reliant les municipalités pour la fourniture des prestations portant sur l'Etat civil), GEONAT (élaboration d'une stratégie nationale dans le domaine de l'information géographique) … sont autant de services qui dénotent d'un contexte de prestation de services électroniques en mutation continue. Ces réalisations sont périodiquement renforcées par des textes juridiques autour de la technologie numérique. A n'en citer, comme échantillon d'une évolution parallèle intégrée, que deux textes de lois d'une valeur stratégique importante. D'une part la loi n°2000-83 du 9 août 2000 portant création de l'Agence Nationale de Certification Electronique qui gère les échanges et le commerce électroniques en instituant le document et la signature électroniques, les services de certification électronique et les transactions commerciales électroniques ; et d'autre part, la proclamation de la loi n°5 de l'année 2004 en date du 03 Février 2004 portant création de l'Agence Nationale de la Sécurité Informatique.

Tout ceci n'est qu'un témoignage d'un cadre favorable pour une migration progressive et prometteuse vers un environnement d'usage intensif de documents électroniques dans tous les secteurs de la société tunisienne. Une nouvelle forme d'expression est désormais mise à la disposition des acteurs de la société tunisienne à travers une typologie de services et de produits numériques d'information qui véhiculent beaucoup de particularisme culturel local.

## 3. Le document numérique : un espace d'expression culturelle et linguistique

Sans vouloir revenir sur la théorie Shanonienne du rapport entre le message et le support, un document numérique est souvent décontextualisé, au moment de sa genèse ou de son usage, de son environnement culturel et linguistique propre. Or, à tout instant de sa vie, il est en contact permanent avec ses créateurs, ses gestionnaires et ses usagers et subit par conséquent les impulsions psycho-cognitives de l'être humain qui conditionne les manières de sa conception, sa maintenance et son interprétation par la machine.

Les documents numériques comportent ainsi un vecteur d'influence linguistique et culturelle imposé par l'intrinsèque liaison qui les associe à leurs deux "congénères", en l'occurrence l'information qu'ils contiennent et les technologies qui servent à les créer ou à les exploiter dans leurs différentes phases de vie. En réalité, tout document à vocation générale ou spécialisée, n'a de valeur réelle que s'il est appréhendé selon les règles définies par son contexte technoculturel environnant. Ces règles constituent un ensemble de paramètres qui traduisent l'efficacité de l'acte de communiquer, l'état des rapports d'usage et d'accoutumance entre information, technologie, modes d'usage et langage de communication.

De ce fait, l'aspect linguistique et culturel a toujours eu une présence évidente dans l'acte de la communication du document électronique. Aujourd'hui il prend de plus en plus d'ampleur, conditionné par les turbulences nationalistes qui émergent comme ripostes à une mondialisation rampante à modèle unique, en l'occurrence occidental, anglo-saxon et plus particulièrement américain. On voit de plus en plus des initiatives nationalistes directement associées à l'identité linguistique et culturelle capable à elle seule de dresser dans l'esprit collectif des individus, des limites identitaires plus tangibles que les frontières politiques. Pour ne rester que dans les limites des pays maghrébins, et particulièrement dans le contexte tunisien qui nous intéresse en premier lieu dans ce document, le recours à un texte juridique imposant l'usage de la langue arabe dans les documents administratifs, l'usage de la langue arabe comme langue première par rapport au français (position et taille) dans les enseignes publicitaires n'est qu'un échantillon de cette importance linguistique qui, pour un certain temps, a bouleversé l'héritage acquis des traditions linguistiques commune tunisiennes dans l'usage de la langue française dans les activités administratives.

Il est donc primordial, comme le confirme A. DANZIN, chef de groupe de réflexion stratégique pour la Commission des Communautés Européennes, que "*le défi linguistique devait être considéré sur le plan économique et social* [et administratif] *comme un phénomène aussi important que le furent, au cours de la décennie 1960, les apparitions de la micro-électronique et les industries des logiciels informatiques* "[**Danzin, 1992**].

L'objet de ce document n'est pas en fait de faire des analyses historiques ou épistémologiques sur le caractère multilingue et multiculturel de la société tunisienne, autant qu'il est une tentative de sonder la profondeur des obstacles techniques auxquels font face les actes de gestion et de communication des documents électroniques dans l'entreprise tunisienne où les utilisateurs cherchent encore une stabilité des normes et une fixation des pratiques dans un référentiel culturel et linguistique conforme à une technologie en place et à une politique administrative en voie de normalisation. Car il faut encore admettre que les acteurs de la société tunisienne (tous niveaux confondus) vivent encore le paradigme de la dualité langagière entre une éducation initiale plutôt francisée et une « contrainte » administrative arabisante. Des études sociolinguistiques démontreraient que la phase de migration vers le « tout arabe » est encore en cours et qu'elle a encore du chemin à faire. L'administration tunisienne est encore à l'âge de transition et d'un bilinguisme relativement « déséquilibré » !

## 3.1. Le bilinguisme arabe/latin

A partir du constat précédent qui dénote d'un taux élevé d'intégration des TIC dans la société tunisienne et de l'importante évolution technologique que connaît l'administration en Tunisie, une première interrogation s'adresserait à la nature des solutions apportées à des strates différentes du modernisme technologique pour surpasser l'obstacle purement linguistique et culturel qui a longuement marqué les premières solutions informatiques basées sur l'exclusivité du codage américain des caractères (ASCII). Face à l'expansion et à la généralisation des systèmes ouverts et distribués, comment le bilinguisme arabe/latin (codage, typographie, morphologie, syntaxe, sémantique et surtout bidirectionnalité de l'écriture), a-t-il trouvé sa voie dans les systèmes électroniques actuels d'information et de communication dans les structures de l'administration et de l'entreprise ? Quel degré d'introduction affiche-t-il ? Et dans quelles conditions de conformité s'opère-t-il avec les usages et les référents socioculturels des utilisateurs ?

Même s'il est communément admis qu'à ce jour, le domaine du multilinguisme numérique a beaucoup évolué grâce aux efforts de beaucoup d'acteurs internationaux, la réponse à cette question générale passe obligatoirement par la réponse à des questions plus pointues : faudrait-il adapter la langue aux critères technologiques des outils de traitement des données ou au contraire faudrait-il que ces derniers se plient aux caractéristiques particulières de chaque langue ? Faudrait-il repenser à la source les structures originales des systèmes informatiques pour y apporter les solutions multilingues ou faudrait-il plutôt accepter l'extension des systèmes monolingues (latins) existants pour en faire des plates-formes bilingues ? Tout un débat international est en cours depuis des années sur ces points clés des systèmes multilingues. Deux nouveaux domaines de recherche appliquée sont depuis lors en effervescence : l'internationalisation (i18n) et la localisation (l10n) des systèmes électroniques dont il sera question dans un chapitre suivant concernant les solutions technologiques proposées.

## 3.2. Caractéristiques et problèmes de la langue arabe

Avant de passer à la formulation de recommandations ou de règles de bonne conduite qu'il serait judicieux d'appliquer pour une harmonisation plus contrôlée des pratiques et des usages des documents numériques entièrement ou partiellement en langue arabe, il serait à mon avis opportun d'introduire les zones de difficulté que la langue arabe affiche pour sa numérisation. Il serait également intéressant de prendre position par rapport aux réalisations déjà enregistrées par l'industrie de la langue pour une transparence linguistique de la langue arabe dans les systèmes électroniques multilingues.

Parmi les caractéristiques qui distinguent la langue arabe des autres langues latines et qui, par conséquent, engendrent des difficultés techniques dans un cadre de bilinguisme arabe-latin, retenons les points suivants :

- Avec ses 28 caractères de base extensibles dans leurs diverses variations à 78 formes graphiques, l'écriture arabe s'inscrit dans son intégralité graphique dans les registres des caractères dits diacritiques (caractères accentués), donc à niveaux de complexité élevés.
- L'aspect cursif de sa transcription (les différents caractères formant le mot sont liés entre eux dans un mode manuscrit ou imprimé) cause souvent des problèmes au niveau de l'affichage, de l'impression et de la reconnaissance optique des caractères . GHOZAL attribue ce handicap à un manque d'industrialisation du au fait que la langue arabe n'a pas été adaptée à l'outil d'impression comme c'est le cas du caractère latin. Ceci est d'ordre à rendre plus difficile les techniques de segmentation du texte arabe dans une opération OCR.
- L'écriture arabe dispose d'une taille irrégulière de corps, ce qui rend difficile l'uniformisation de la chasse et de la hauteur des caractères.
- Le système de vocalisation est basé sur des diacritiques supplémentaires (shakl, tanouin et shadda) utilisés presque exclusivement pour les textes sacrés et l'apprentissage de la langue.
- Les caractères peuvent avoir plusieurs formes en fonction de la position qu'ils occupent dans le mot : au début, au centre, à la fin ou isolé. La lettre peut avoir des variantes graphiques selon sa position isolée, initiale, médiane ou finale. Cette opération s'effectue grâce à la procédure d'analyse contextuelle par laquelle les lettres arabes acquièrent des formes différentes suivant leur position dans le mot, et suivant les lettres environnantes.

- L'analyse contextuelle dans la langue arabe est encore plus complexe. Sa difficulté est la présence de voyelles et autres signes diacritiques qui se placent au-dessus ou au-dessous des lettres et qui ne doivent pas affecter la contextualité. Pour accélérer le calcul des formes contextuelles, tous les systèmes d'exploitation l'ont intégré dans leurs couches logicielles les plus profondes ; ainsi la contextualisation est indépendante du logiciel utilisé. Cette intégration a des conséquences fâcheuses : l'utilisateur final ne peut y accéder, et ne peut modifier les règles contextuelles. Ainsi, par exemple, il est impossible d'ajouter une lettre arabe parce que cette lettre aurait à interagir avec les autres ce qui impliquerait une modification des règles contextuelles (alors que dans le cas des autres écritures, il suffit de dessiner un caractère pour pouvoir l'utiliser tout de suite dans un texte).
- La possibilité de superposition verticale de caractères (à l'instar de la fusion horizontale de certains caractères latins comme dans ,  , ), rend difficile la reconnaissance optique des caractères, l'affichage et l'impression.
- L'écriture et la lecture de l'arabe s'effectuent de gauche à droite. Ceci est d'ordre à introduire des algorithmes supplémentaires pour gérer le changement du sens d'affichage ou d'impression entre l'arabe et le latin dans des applications bilingues où des actions de césure et de bris de texte s'applqient.
- La langue arabe gère deux types de virgules codées différemment : une numérique identique à la virgule latine et une textuelle inversée et orientée à droite.
- Les formes minuscule et majuscule des caractères sont inexistantes.

## 3.3. Codage des caractères, langues et normalisation : l'effet Unicode

La technologie, dans son essence et sa genèse, est un vecteur de dominance culturelle bien connue, en l'occurrence le monde à dominance linguistique latine et plus particulièrement anglo-saxonne. Par les voies de la mondialisation, l'outil et ses mécanismes de fonctionnement se sont fait appropriés par la communauté internationale. Les produits d'information qui en découlent ne sauraient échapper à cette empreinte qui marque en quelque sorte la présence, quoi que parfois subliminale, de la langue de l'outil. Il suffit de voir les zones systèmes (protocoles, URI etc.) restés exclusivement limités à la codification ASCII pour comprendre l'approche subliminale de la domination linguistique anglo-saxonne. Moins subliminales étaient les obstacles qui relèvent des domaines de la codification électronique des données, de la normalisation des procédés de transcription et d'échange.

En effet, la situation générale démontre aujourd'hui, que tout fonctionne majoritairement dans un mode dominé par la langue de la ressource et/ou du serveur d'information. Dans des systèmes latins hétérogènes (français, anglais, espagnol ...), la solution du multilinguisme s'est faite possible grâce à une extension banale des codes de représentation donnant lieu à des jeux de caractères codés sur 8 et 16 bits (ISO 8859 et Unicode) et de l'interposition d'artefacts qui relèvent du domaine de l'ingénierie linguistique (i.e. TAO).

Cependant, la présence des langues non latines (i.e. arabe, hébreux, chinois, cyrillique ...) bien qu'aujourd'hui très réconfortées par l'avènement d'Unicode, reste encore exceptionnelle dans les environnements applicatifs spécifiques comme les noms de domaines, les modes opératoires des protocoles de transferts de données etc.

L'avènement d'Internet allait renforcer en quelque sorte la dimension multilingue des ressources et des services. Devenue de nos jours l'alternative par excellence pour les utilisateurs des systèmes d'information distribués, et le modèle par excellence pour les entreprises dans la mise en place de leurs Intranets et Extranets, son architecture client-serveur du type WWW a pu rendre possible l'accès à l'information sans obstruction linguistique aucune. Les développements d'interfaces multilingues, dynamiques et intelligentes avec des potentialités de reconnaissance de caractères non latins (arabe, japonais, russe ...) prolifèrent dans l'univers Internet. Ce stade d'évolution et cet état d'universalité n'ont pu devenir réels sans la contribution d'un effort de normalisation qui a accompagné la conception des systèmes bilingues depuis leur début.

La langue arabe a pris sa part dans cet effort de normalisation bien que des controverses encore en vigueur perturbent sa fixation sur des assises normatives définitives. Sans nous approfondir dans les détails de l'historique de la normalisation du caractère arabe, largement gérée par l'ASMO (Arab Standards and Metrology Organisation) et l'ISO, l'intérêt aujourd'hui est porté vers les nouveaux protocoles, langages et systèmes qui adhèrent pleinement à l'effort de l'internationalisation des systèmes automatisés de l'information.

L'un de ces systèmes est sans doute le protocole de transfert des données (*http, hypertext transfert protocol*) et le langage dans lequel les ressources sur Internet sont décrites (*HTML, HyperText Markup Language*). Dans ses nouvelles versions, le langage HTML 4.01 (recommandations W3C) a évolué vers une reconnaissance de la langue arabe selon un encodage et un balisage spécifique. L'internationalisation des langages HTML et XHTML a introduit beaucoup de nouvelles propriétés visant à faciliter l'usage de l'arabe dans le Web. Ces facilités sont basées essentiellement sur Unicode et sont intégrées dans la version HTML4.01 et XHTML1.0. Unicode définit pour chaque caractère le sens de son écriture (LTR ou RTL) selon un algorithme de bidirectionnalité pour gérer les bris de textes mixtes (latins et arabes).

## 3.4. Les usages des documents numériques : l'empreinte socioculturelle

La pratique actuelle de la langue arabe dans un contexte informel oral ou dans un environnement formel écrit, a subi certaines modifications engendrées principalement des habitudes d'usages qui, à travers le temps, se sont enracinées dans les pratiques rendant difficiles toutes tentatives d'harmonisation et de conformité. Ces habitudes divergentes sont actuellement visibles au niveau de :

- des manières d'usage des chiffres et des dates dans les documents officiels. Aucune norme n'établit jusqu'aujourd'hui une règle de conduite uniforme pour tout le monde arabe afin d'utiliser un modèle unique de chiffrement. Les écarts s'observent essentiellement au niveau de la directionnalité des valeurs de pourcentage et des chiffres à valeurs décimales. Il est encore courant de trouver des érudits mettant la décimale à gauche d'un chiffre ou illustrant le signe du pourcentage à gauche de la valeur correspondante. C'est là l'une des polémiques engendrées par l'undirectionalité gauche-droite des chiffres dans les systèmes multilingues même à sens d'écriture originel opposé comme l'arabe.

- de la graphie utilisée pour représenter les chiffres (hindous ou arabes). Concentré dans la partie orientale du monde arabe (à partir de la Libye vers le Moyen Orient), l'usage des chiffres hindous est entièrement rejeté par les maghrébins (Tunisie, Algérie, Maroc, Mauritanie) qui s'attachent pour des raisons culturelles aux chiffres arabes. En orient, il est par contre possible d'aller contre la règle et de faire usage des deux systèmes de numérotation.

- de la présentation de la vocalisation avec le caractère arabe de base. La tendance est de voir aujourd'hui les diacritiques arabes s'afficher dans les textes de la littérature courante (journaux, livres, magazines) alors qu'auparavant elle se limitait uniquement aux textes sacrés et aux manuels d'apprentissage de la langue.

- de la combinaison des caractères sous forme de ligature et par superposition verticale de deux ou trois caractères au plus. Ceci ne figurait pas parmi les pratiques calligraphiques arabes classiques.

Parmi les conséquences à cette diversité dans les usages, les documents numériques se trouvent affectés dans leurs missions d'information et de communication au sein d'une communauté large qui cherche à confirmer son identité culturelle et linguistique à travers les réseaux pour les besoins d'échange, de partage et de diffusion de son patrimoine culturel.

## 4. Le document numérique multilingue : un besoin d'universalité dans les solutions techniques, culturelles et linguistiques

S'il était question dans le chapitre précédent de faire le tour d'horizon autour des caractéristiques de la langue arabe et des particularités de sa numérisation, l'objectif est d'arriver à contourner les points d'achoppement et à entreprendre les mesures qui pourraient faciliter la progression vers une harmonisation du contexte de l'édition des documents numériques arabes dans la société tunisienne, leur gestion et leur exploitation selon des principes d'uniformité maximale. L'aboutissement final reste la progression vers un système de gestion de documents administratifs ou de réservoirs de documents d'archives qui seraient capables de favoriser des services à valeurs ajoutées comme la recherche, l'échange, le workflow et le datamining.

L'aboutissement à cet état de fait ne saurait concerner le document numérique à lui seul. Son environnement de genèse, de traitement, de manipulation et de conservation sont autant concernés par ces mesures d'uniformisation.

## 4.1. L'internationalisation des systèmes (i18n)

Par internationalisation, il est souvent fait référence, sur un plan technique, aux solutions logicielles et applicatives dont les caractéristiques et les codes ne sont pas fondés sur un seul paramètre de lieu. Ces solutions spécifient en réalité les caractéristiques culturelles des groupes diversifiés d'utilisateurs comme leurs systèmes métriques, leurs calendriers, leurs systèmes monétaires etc. Il s'agit en fait de ressources informatiques (fichiers systèmes connus sous le nom de locales)**[CASTYDE, 2003]** prévues par les concepteurs pour donner davantage de chances d'écoulement sur les marchés mondiaux. Prévoir une panoplie de paramétrages linguistiques et culturels pour un large éventail de communautés linguistiques et culturelles dans le monde est une garantie de vente. Sauf que les degrés de perfection de l'I18n varient en fonction de plusieurs critères. Les exigences de qualité des publics cibles et les alternatives qui leurs sont données influent sur la qualité des produits internationalisés. Je cite à titre d'exemple des produits comme ceux de Microsoft. Par un souci d'universalité, les produits Microsoft ont été tous traduits vers toutes les langues à valeur commerciale évidente. La langue arabe en a eu sa part de couverture sous formes de produits logiciels bilingues entre l'arabe comme langue de base associée à l'anglais ou au français en fonction de la région concernée (français pour l'Afrique du nord, anglais pour le moyen orient). Cette dualité linguistique était par contre arrêtée avec les versions 95 et 98 de Windows qui allaient se limiter à la source (des mises à jours linguistiques ont été ensuite bricolées) aux seuls produits multilingues arabe-anglais. Parmi les raisons préconisées pour cette rupture de dualité linguistique, était cité le déséquilibre du potentiel économique entre le moyen orient et le Maghreb. Pour des raisons d'économie d'échelle, la partie de l'Afrique du Nord, accusée, à raison ou à tors de piratage intensif, a été abandonnée par Microsoft aux produits à soubassements anglophones. L'arrivée massive d'Unicode avec Windows 2000 et XP allait remettre la pendule à l'heure.

## 4.2. La localisation (l10n)

Si l'internationalisation est généralement appréhendée comme une procédure d'ingénierie informatique, la localisation est souvent conçue comme un acte de traduction ou d'adaptation d'une application aux particularités locales d'un utilisateur ou d'une communautés d'utilisateurs. La localisation travaille généralement sur l'adaptation de l'interface (ergonomie) ou de la documentation technique aux besoins linguistiques et culturels des utilisateurs.

Par contre, la localisation ne consiste pas uniquement à choisir au moment de l'affichage les caractères appropriés. Il est nécessaire de bien choisir le bon jeu de caractères, d'afficher les dates, les heures, les montants monétaires et les nombres dans le format accepté par le public cible, d'adopter les conventions locales de tri des chaînes de caractères, et même d'interpréter correctement les saisies des utilisateurs.

Plus la variété des publics cibles est grande, surtout si elle inclut des personnes écrivant de droite à gauche et/ou de haut en bas, toute la structure de l'application en cours de localisation doit être reconsidéré à chaque fois.

Dans tout système d'information, les utilisateurs de logiciels s'attendent à trouver les éléments propres de leurs cultures et de leurs langues. Entre francophones et anglophones, l'ordre de la structure dune date varie du quantième, du mois, puis de l'année pour le francophone à l'opposé pour l'anglo-saxon. Le symbole monétaire, le calendrier, le système métrique etc. sont autant de particularités que l'application devrait permettre à l'utilisateur pour façonner son document numérique sur des assises culturelles et linguistiques qui lui sont propres. « *Tout ceci doit naturellement se faire de manière transparente pour l'utilisateur, qui ne doit à aucun moment percevoir les difficultés de la localisation. Au contraire, le* [système] *doit détecter sa culture et s'y adapter au mieux et automatiquement, sans demander d'action de sa part* »**[Le Roy]**

A l'exception de quelques grands gourous du logiciel comme Microsoft, relativement peu de solutions logicielles et de systèmes d'exploitation prévoient l'usage courant de ces paramètres de localisation dans une approche basée sur l'analyse des usages.

## 4.3. Arabisation et bilinguisme

La langue arabe, encore débutante dans le domaine du bilinguisme systémique, à été l'objet, comme le signale la littérature spécialisée, de plusieurs essais d'intégration ou de conversion vers des systèmes informatiques multilingues. Les projets de systèmes bilingues arabe/latin abondent dans la littérature scientifique. Seulement, pour des raisons de divergences d'approches conceptuelles des dits systèmes, ni les uns adaptant la langue à l'outil informatique (approche exogène) ni les autres effectuant la procédure inverse (approche endogène), ne parvinrent à établir une règle définitive. Il a fallu attendre l'intervention des grands producteurs de matériels et des grands concepteurs de logiciels pour voir les grandes firmes comme Microsoft, Alis Technologies etc. imposer des normes et des standards d'arabisation aux niveaux des matériels et des logiciels issues de leurs propres études de milieu et conformément à leurs propres perceptions du bilinguisme.

L'arabisation des systèmes informatiques monopostes ou multiutilisateurs concerne en effet, les deux aspects fondamentaux des composantes matérielles d'une part et des programmes incluant les langages de programmation, les utilitaires, les progiciels d'application et les systèmes opératoires de l'autre.

En fait, le principe d'arabisation signifie implicitement un bilinguisme arabe/latin basé sur le principe de la transparence pour rendre la machine contrôlable simultanément et de façon transparente par les deux ou plusieurs langues utilisées.

En définitive, l'arabisation est plutôt orientée vers l'accomplissement des objectifs suivants :

* rendre l'usage de la langue arabe en informatique aussi simple et efficace que les langues latines,
* concevoir des applications indépendantes de la langue de dialogue,
* gérer des données en plusieurs langues simultanément,
* manipuler, grâce au principe de la transparence, des données arabes et latines par des applications latines standards sans y apporter une quelconque modification,
* arabiser toute application standard par une simple traduction du dialogue opérateur,
* concevoir des applications intégralement indépendantes de la langue de dialogue choisie par l'utilisateur et intégrant simultanément les différentes langues.

L'accomplissement de ces objectifs a toujours été confronté à des problèmes liés à l'intégration des fonctions d'arabisation que ce soit au niveau du système d'exploitation ou au niveau des dispositifs d'entrée-sorties de présentation.

Parmi ces problèmes, certains ont leur origine dans la langue arabe elle même; d'autres sont issues de la nécessité de mixer la langue arabe avec d'autres langues d'origines latines, d'autres proviennent directement du champs d'application des normes et des codes de représentation.

## 5. Systèmes ouverts et distribués : le besoin des normes et des standards

Pour revenir à l'objectif central de ce document, en l'occurrence l'analyse de l'environnement technologique et des usages dans une optique d'uniformisation des pratiques chez les concepteurs des documents numériques, une voie incontournable est déjà tracée même à échelle internationale : la voie de la standardisation et de la normalisation.

Tout document électronique est aujourd'hui susceptible d'être transmis, échangé, converti, édité, stocké, indexé, récupéré etc. dans des environnements de travail multiples et variés. On prône désormais vers les systèmes ouverts et distribués basés sur les architectures client-serveur et les systèmes intelligents. Les entreprises convergent vers les modèles dynamiques de gestion des ressources d'information, de conservation de fonds d'archives … tous basés sur un principe essentiel, celui de la fluidité des données, de la traçabilité des transactions, du partage des ressources et de la capitalisation de la connaissance. Les réseaux sont là pour répondre à ces exigences dont les objectifs centraux tournent autour de l'économie des ressources et du temps, de la fiabilité et de la pertinence des services rendus.

Or, pour atteindre ces objectifs, un minimum de conformité est nécessaire pour faire face à l'hétérogénéité des applications informatiques, des systèmes d'exploitation des machines et des marques de fabrication des ressources matérielles. Le domaine des standards et des normes est ainsi apparu depuis quelques années pour couvrir tous les aspects du fonctionnement de la société savante y compris le domaine des documents numériques et des archives. Les exemples de la description normalisée des archives ISAD(G) et du schéma des métadonnées archivistiques EAD en sont la confirmation.

L'univers des normes et des standards est très vaste. Notre intérêt portera dans ce document sur deux constituantes essentielles des documents numériques, à savoir, son contenu d'information et sa structure physique comme support d'information et moyen de communication. Trois concepts clés seront abordés comme recommandations d'usage pour une optimisation du contrôle des gisements de documents émis et exploités par les services producteurs de documents numériques : les métadonnées, les structures décentralisée et les réseaux sémantiques des documents numériques.

## 5.1. Les métadonnées : formes et usages

Devant l'accroissement vertigineux de la masse des documents générés, échangés et conservés par les différents acteurs de la société de l'information, le concept des métadonnées est réapparu comme solution idéale pour optimiser le contrôle de cette masse d'information et son exploitation pour des affinités multiples.

L'exemple classique toujours utilisé pour décrire les métadonnées, c'est la fiche cartonnée d'un catalogue de bibliothèque. Cette fiche, elle même normalisée, permet de synthétiser le contenu et la forme du document sans besoin d'y avoir accès. En somme, toutes données descriptives du genre résumé, mots clés ou tout ce qui peut être un substitut au document original sont assimilées à des métadonnées.

Dans le contexte informatisé, les métadonnées sont des données qui renseignent sur des objets numériques qui peuvent prendre plusieurs formes : des objets textuels, des images, des fichiers son, des séquences vidéos. En décrivant les attributs et le contenu de ces objets, les métadonnées deviennent utiles au repérage et à la restitution de ces documents noyés dans une masse volumineuse d'information. Elles servent aussi à la gestion de l'information selon le schéma classique de la chaîne documentaire : description, accès et conservation des données.

Nous n'entrerons pas ici dans l'exposition des chiffres astronomiques des masses documentaires sur Internet pour justifier l'usage des métadonnées à des fins de contrôle, de gestion et de restitution, nous extrapolons simplement le modèle Internet sur l'environnement des entreprises et des administrations pour retrouver le même modèle opérationnel des métadonnées dans la gestion des ressources d'information au sein des Intranet à des affinités de procédures et de fonctionnalités diverses comme le Workflow et le datamining. Mais l'entreprise ou l'administration n'est pas uniquement une masse d'informations régies en Intranet. On estime que 83% des documents numériques sur Internet sont d'ordre commercial, ce qui dénote de la contribution des entreprises et des structures apparentées dans la création et l'édition des documents numériques diffusés sur les réseaux.

Plusieurs raisons justifient le recours à des métadonnées dans ce contexte à la fois riche et pléthorique des entreprises. L'accessibilité et l'utilisation fluide des documents électroniques en sont quelques uns justifiant la qualité et la compétitivité économique et industrielle. Des outils de recherche basés sur l'intelligence des systèmes électroniques mis en place, doivent permettre une dynamique de datamining optimale et une jonction synchronisée avec le front office de recherche et d'accès aux ressources. Par le recours aux données descriptives, la re-cherche d'information peut s'effectuer dans plusieurs collections de documents et l'interopérabilité entre plusieurs structures. L'Initiative des Archives Ouvertes (OAI) est aujourd'hui un exemple concret.

La méthode à suivre pour faire usage des métadonnées consiste à associer des éléments descriptifs le plus tôt possible aux ressources numériques créées vu qu'elles peuvent également servir pour la production d'autres ressources électroniques.

Deux approches génériques sont prévisibles dans ce sens : les métadonnées pour les documents crées à la source par des procédés numériques et les métadonnées ajoutées aux documents rétrospectifs numérisés par scannérisation.

Pour le premier type de procédé, les éditeurs de nouvelles génération on tendance à utiliser automatiquement des éléments de métadonnées standards. Depuis les solutions bureautiques courantes comme Microsoft Office, passant par les solutions GED d'Acrobat et arrivant aux solutions HTML et XML, les applications et les éditeurs de création de documents numériques donnent la possibilité de paramétrer des éléments personnalisés de métadonnées en plus des formats propriétaires prévus par l'application elle même (Fig.1). Or les analyses ont toujours démontré qu'il y a une négligence quasi générale à ces champs de description malgré leur importance dans les processus d'indexation de ces ressources dans leurs systèmes d'information natifs.

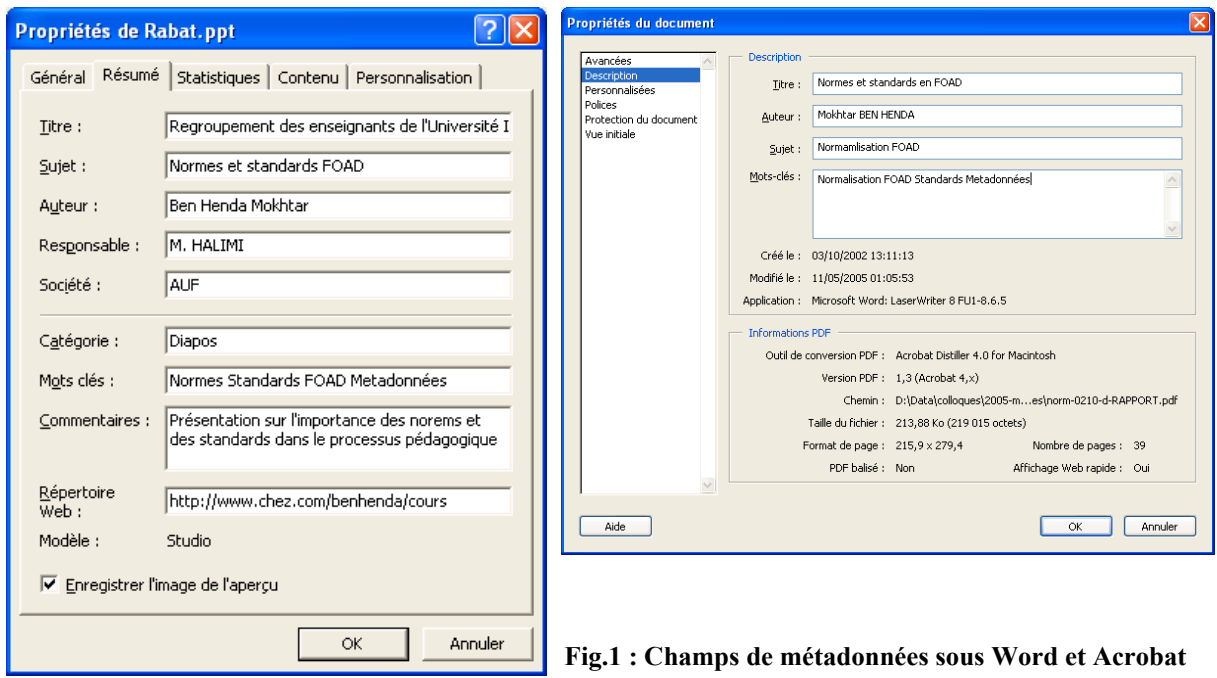

Fig.1 : Champs de métadonnées sous Word et Acrobat

Pour le deuxième type de méthode, la numérisation par scannérisation des documents papiers résulte toujours en un support d'information graphique. Les images des textes obtenus par la numérisation sont aussi susceptibles d'être référencées par des éléments de métadonnées que les applications GED ou GEIDE permettent désormais d'adjoindre aux copies électroniques des supports papier. Une phase avancée d'océarisation (reconnaissance optique des caractères) optimiserait les possibilités d'indexation et de recherche des documents en question. La suite Acrobat excelle en la matière avec ses solutions OCR et d'indexation intégrées.

### 5.1.1. Quelques modèles de métadonnées

Depuis l'apparition des métadonnées comme ressources vitales aux systèmes d'information électronique, les modèles de leurs conceptions et de leur intégration dans les documents numériques prolifèrent chaque jour au point de créer une divergence qui rend l'objectif de contrôler la masse des documents électroniques difficile à atteindre. Les voies de la standardisation se sont alors élevées pour harmoniser un tant soit peu toutes ces initiatives et leur créer des points de convergence capables de les amener à des possibilités d'échange et de conversion. Chaque domaine s'est doté de son propre modèle et chaque consortium s'est barricadé derrière son schéma propre. On recense aujourd'hui des centaines de schémas de métadonnées tous domaines confondus. Les plus connus ayant rapport avec les documents numériques sont particulièrement les standards des objets pédagogiques (LOM , SCORM, NORMETIC etc.). Or ces modèles sont eux-mêmes une amélioration de modèles plus anciens comme le non moins fameux Dublin Core, l'un des standard de métadonnées le plus répandu et le plus avancé pour la description des ressources sur Internet.

Beaucoup plus récente, la norme RDF (Resource Description Framework) facilite davantage le traitement des métadonnées en assurant un degré optimal d'interopérabilité entre les applications grâce à son utilité polyvalente pour la découverte de ressources, le catalogage, l'évaluation du contenu, la gestion des droits d'auteur.

Le domaine des archives est l'un des domaines les plus récent à s'intéresser aux métadonnées normalisées. Le Conseil International de Archives travaille déjà sur l'application d'un modèle unifié de schéma de métadonnées archivistique basé sur le modèle EAD (Electronic Archival Description) basé sur la structure descriptive de la norme ISAD(G).

### 5.1.2. Les Ontologies : un vocabulaire normalisé pour la description des documents électroniques

L'apparition de tous ces modèles et de ces schémas de métadonnées n'a pourtant pas résolu le problème de la complexité des systèmes d'information numérique bien qu'il suffit de remonter le cours de l'histoire de la gestion des documents d'entreprise pour voir qu'elle était l'apanage des informaticiens qui ont mis l'accent sur les aspects techniques touchant la structuration et l'indexation des documents ainsi que les formats d'échange et les réseaux qui les véhiculent. Un problème majeur reste pourtant toujours d'actualité. Avec l'apparition de concepts comme la gestion des connaissances, le besoin d'une indexation sémantique s'est vite ressenti pour gérer les documents selon leurs contenus.

En effet, si l'on parvient à unifier, ou du moins rapprocher, la structure des métadonnées pour la description des ressources numériques, comment peut-on assurer la similitude, ou le rapprochement, dans le sens de la description ? En d'autres termes, si l'aspect structurel de la description est relativement harmonisé, l'aspect sémantique requiert un intérêt particulier car c'est à travers lui que s'évalueront les accès et les échanges des données numériques. Ceci nous remet dans le contexte classique et limité des thésaurus, des lexiques d'entreprise et des langages contrôlés d'indexation avec tout ce que cela représente au niveau de la gestion des synonymes et de l'unification des concepts. Or, ces outils sont désormais insuffisants pour couvrir la richesse des contenus des documents techniques et les performances de traitement énormes des dispositifs technologiques d'aujourd'hui. On parle désormais de taxonomies et d'ontologies basées sur des catégories de données et des liens qui les unissent dans des réseaux sémantiques.

Les ontologies sont des systèmes de représentation relationnelle associant niveaux lexical (termes) et conceptuel (concepts) qui permettent de raisonner sur un domaine. Elles touchent plusieurs acteurs qui gèrent la documentation d'entreprise depuis les informaticiens, aux documentalistes, terminologues et traducteurs. Elles facilitent les opérations de recherche d'information, d'extraction des données, d'élaboration de métadonnées, de veille scientifique et d'aide à la traduction...

## 5.2. Les documents structurés

La richesse qu'a connue le domaine des métadonnées a été engendrée en partie par une évolution dans la structure des documents numériques. Après une période de structuration en unités homogènes de documents séquentiels, l'hypertexte d'une part et les métadonnées de l'autre ont donné lieu à une nouvelle catégorie de documents numériques virtuels et éclatés. En effet, en même temps que les textes deviennent plus riches, d'autres types de médias s'y ajoutent. L'intégration de ressources multimédia rejoint très vite l'univers des documents numériques structurés. Des nouvelles formes de gestion des ces ressources sont devenus indispensables pour suivre cette évolution dans sa décentralisation, sa modularité, son extensibilité et sa simplicité. Un mot maître dans ce sens : XML (Extensible Markup Language).

XML est un langage de représentation des données et de documents structurés. Sa particularité est de n'avoir aucune sémantique qui pourrait l'empêcher de s'appliquer à tout genre d'application à des fins de représentation et d'organisation de tout type de donnée ou de document. Ouvert et extensible, il peut décrire chaque type de données dans des formats de documents adaptés à tout contexte.

Les performances de XML dépassent la simple représentation de documents équivalents à des pages Web habituelles. Il offre une élasticité de structuration qui permet à chaque secteur d'activité de définir le format de document qui lui convient. Il permet aussi de représenter des structures inhabituelles dans les documents conventionnels, comme la structure temporelle des documents multimédia. Grâce à Unicode, il peut représenter des documents multilingues.

Or, bien que XML dispose d'une grande capacité de description des documents, il se limite à décrire la structure en délaissant la description du sens à des outils complémentaires comme les métadonnées dans les DTD (Data Type Definition) et les schémas RDF.

Avec tous les compléments qui lui sont associés pour gérer les styles et les aspects graphiques des documents (CSS, XSL), ou pour appliquer des modèles de schémas de structures (Schémas XML) ou appliquer des langage d'accès aux bases de données (XQuery), ou pour exprimer des transformation capable de dériver automatiquement plusieurs documents différents, dans différents formats, à partir d'une seule source XML (XSLT), le langage XML s'érige aujourd'hui comme la solution la plus appropriée pour la gestion des documents numériques sur les réseaux d'entreprises.

**Fig.2 :** *Le modèle XML* Un document XML a besoin d'une DTD (incorporée ou reliée) pour identifier les éléments et les attributs à trai-

Le fichier «annuaire.dtd» déclare les éléments sémantiques à traiter.

```
<?xml version="1.0" encoding="ISO-8859-1"?>
<!ELEMENT annnuaire (personne*)>
<!ELEMENT personne (nom,prenom,email+)>
<!ATTLIST personne type (étudiant | professeur | chanteur | musicien) "étudiant">
<!ELEMENT nom (#PCDATA)>
<!ELEMENT prenom (#PCDATA)>
<!ELEMENT email (#PCDATA)>
```

Un balisage sémantique XML sur un modèle DTD externe.

```
<?xml version="1.0" encoding="ISO-8859-1"?>
<!DOCTYPE annuaire SYSTEM "annuaire.dtd">
<annuaire>
    <personne type="étudiant">
        <nom>HEUTE</nom>
        <prenom>Thomas</prenom>
        <email>webmaster@xmlfacile.com</email>
    </personne>
    <personne type="chanteur">
        <nom>CANTAT</nom>
        <prenom>Bertrand</prenom>
        <email>noir@desir.fr</email>
    </personne>
</annuaire>
```

## 5.3. Les réseaux sémantiques

L'une des évolutions engendrées par l'usage des métadonnées est sans doute la notion des réseaux sémantiques prioritairement attribuée au Web. Dans le Web d'antan, seul un être humain pourrait analyser le contenu intellectuel d'une information disponible sur le Web. Les machines n'avaient pas les capacités d'analyse nécessaires pour interpréter. Elles servaient uniquement à localiser  transférer et appliquer des mises en forme et des présentations. Dans le Web sémantique, l'information a un sens explicite hérité de l'usage de deux concepts :

1. la structuration plus riche, plus agencée et plus rigoureuse de l'information elle-même,
2. les métadonnées servant à décrire rigoureusement l'information principale. Des calculs sur les métadonnées permettent d'inférer des propriétés sur les données qu'elles décrivent.

La prolifération de documents numériques normalisés et transmissibles selon un sens et un contexte d'usage conduisent inexorablement à parler de *"réservoir de documents numériques"* ou d' *« entrepôts de ressources numériques »*. On retrouve dans cette définition tout le sens d'un lieu unique et virtuel qui regroupe diverses fonctions ayant trait à l'utilisation sociale des documents numériques :

- un réservoir de documents, de préférence mis à jour de façon dynamique et accessible à tous ;
- un lieu de confrontation de l'utilisateur avec le document numérique dans une situation qui permet de lire un document, de le retrouver dans la masse des documents, de le citer pour le relire et le critiquer ;
- un lieu de conservation des documents pour des délais qui dépassent la vie des individus ;
- un lieu d'organisation des connaissances par la classification et le référencement des documents. Le fait qu'ils permettent l'accès aux documents par leurs contenus, les outils de recherche dans les réservoirs des documents numériques modernes dépasse leur rôle d'outil de gestion pour devenir un instrument de valorisation des contenus ;
- un lieu où sont rendus des services au public : aide aux chercheurs, conseils aux décideurs, informations locales et internationales sur les organismes et les associations ;
- un lieu de sauvegarde du patrimoine culturel et linguistique et de sauvegarde de la mémoire collective.

Comment retrouver tous ces aspects des systèmes d'information dans l'univers des documents numériques en réseau ? C'est le pari auquel sont confrontés l'administration publique et les acteurs privés indécis entre les pressions des droits à l'information et des contraintes sécuritaire, les attentes du public pour un accès gratuit et généraliste à toute l'information, et les contraintes techniques et organisationnelles qui se préservent le droit de la confidentialité.

## 5.4. Les valeurs culturelles et linguistique

Si nous avons pu nous étaler sur des considérations techniques afférentes aux structures et contenus sémantiques des documents numériques de façon générale sans trop converger vers les particularités culturelles et linguistiques, c'est que ces deux aspects se jouent en réalité sur des dimensions relativement disjointes de ces aspects. Les structures XML comme les ontologies ou les réseaux sémantiques sont une affaire d'ingénierie linguistique qui se greffe ou s'associe à ces considérations techniques pour orienter une solution ou une autre vers une dimension culturelle ou linguistique particulière.

Déjà, Unicode est devenu le jeu de caractères codés implicites pour les structures en XML. Dans le prologue des documents XML, la déclaration de la technique d'encodage fait référence à des normes de codages multi-octets.

```
<?xml          version="1.0"          encoding="utf-8"?>
<!DOCTYPE   html   PUBLIC   "-//W3C//DTD   XHTML   1.0   Transi-
tional//EN"
"http://www.w3.org/TR/xhtml1/DTD/xhtml1-transitional.dtd">
```

Les métadonnées, les ontologies et les réseaux sémantiques, par contre, sont encore l'apanage des langues et cultures latines et germaniques dominantes. Tout l'intérêt est de développer ce genre d'outils et de processus pour préserver les particularités culturelles et linguistique des entités éditrices de documents numériques ou celles des utilisateurs potentiels de ces ressources.

## *6. Recommandations*

Pour être plus concret dans cette approche descriptive du contexte global des documents numériques en entreprise, il serait peut être temps de proposer une série de recommandations dont l'objectif serait d'orienter vers une optimisation des acquis de la numérisation pour harmoniser en amont du processus archivistique ou de gestion des documents administratifs, une démarche capable d'unifier les processus de création, de gestion et de conservation des documents numériques. L'intérêt sera porté, en fonction de l'optique de ce document, vers les consideration culturelles et linguistiques locales qui caractérisent une large panoplie des documents rétrospectifs et courants de l'activité administrative tunisienne. Trois grandes catégories de recommandations sont proposées ci-après.

### 6.1. Codage universel

```
<?xml version="1.0" enco-
ding="utf-8"?>
<!DOCTYPE html PUBLIC "-
//W3C//DTD XHTML 1.0 Transi-
tional//EN"
"http://www.w3.org/TR/xhtml1/D
TD/xhtml1-transitional.dtd">
<html dir="rtl" lang="ar"
xml:lang="ar">
<head>
 <meta http-equiv="Content-
Type" content="text/html;
charset=utf-8" />
<title>        <title>
```

L'encodage est déclaré par une balise <**meta**>. La bonne pratique recommande d'inclure toujours cette balise, même si l'encodage peut déjà avoir été spécifié dans le paramètre http(Content-Type) ou l'entête XML (cas de XHTML). En effet, cela garantit que l'encodage est toujours déclaré au cas où la page est enregistré sur disque local ou a migré vers un autre site, non configuré pour servir Content-Type.

La langue est spécifiée dans la balise <**html**> par l'attribut HTML **lang** (ou XHTML **xml:lang**). Ici la langue arabe **ar**, est la langue première de tout le document. Les valeurs de

L'une des conséquences de l'usage des systèmes ouverts et distribués, est sans doute l'augmentation de la couverture géographique des réseaux d'information. Des populations de plus en plus variées accèdent au réseau. La multiplicité des langues devient primordiale. Tous les systèmes d'écriture doivent être utilisables, séparément ou ensemble, dans de véritables documents multilingues. En plus de l'enrichissement sémantique évoqué plus haut, il devient nécessaire d'adopter des représentations universelles des contenus textuels. Il est donc de l'intérêt de tous les acteurs actuels ou futurs des documents numériques que ceux-ci soient conçus dans un jeu de caractères universels capable d'être lu et déchiffré partout dans le monde sous toute forme d'application capable de comprendre le jeu de caractères universels. Le standard Unicode est recommandé dans ce cas de figure dans l'une de ses sa forme complète (UCS) ou réduites (UTF).

Le codage linguistique ne se contente pas de spécifier le jeu de caractères utilisé. Il doit insister sur la spécification des codes des langues et de la directionnalité de l'écriture.

langue à deux lettres ar fr ... sont définies par une norme ISO. L'attribut lang est là pour permettre aux agents clients, le cas échéant, de compléter le rendu d'un document avec des spécificités propre à une langue. Par exemple pour l'arabe, utiliser toujours une police stylisée.

### 6.2. Optimisation des métadonnées

Tout document numérique est désormais associé à une fiche descriptive qui reproduit ses métadonnées dans le format utilisé. La recommandation est une forte incitation à décrire l'information d'une façon structurée et rigoureuse, au sein de toute application utilisée (Office) ou indépendamment d'un outil particulier (XML). Pour permettre aux nouvelles applications et aux nouveaux appareils de tirer profit de l'information que nous publions, il faut respecter ces formats. Sinon, l'information produite sur mesure pour l'outil à la mode au moment de la publication deviendra inaccessible et inexploitable dès que cet outil aura disparu.

Heureusement, en même temps que les nouveaux formats apparaissent les outils qui les supportent deviennent disponibles, en particu-lier les outils de création et de mise à jour. Les bons outils XML sont capables de garantir que les documents qu'ils produisent sont conformes aux schémas choisis. Il devient aussi plus facile de vérifier la qualité de ces outils avec les différents services de validation et les suites de tests disponibles.

On dispose maintenant de formats ouverts, c'est à dire complètement documentés, mis en oeuvre dans des applications différentes qui proviennent de plusieurs sources. Cela garantit la pérennité de l'information et donne les meilleures chances de pouvoir l'exploiter pendant longtemps, même avec des applications qui ne sont pas disponibles aujourd'hui.

Ces formats ouverts sont eux-mêmes définis de façon ouverte et coopérative, pour la plupart par le W3C (World Wide Web Consortium). Ils bénéficient de l'expérience des experts du domaine et des contributions des principaux industriels impliqués. Ils garantissent la pérennité de nos données et leur intégration dans le Web de demain.

L'usage de métadonnées multilingues nécessiterait obligatoirement le développement d'une terminologie unifiée. En d'autres termes, il serait opportun de développer des ontologies spécialisées multilingues qui répondraient d'une part au champ spécifiques d'activité de l'entreprise et aux caractéristiques propres aux utilisateurs de ces ressources numériques.

### 6.3. « Scannérisation Océarisée »

Une bonne partie des ressources numériques dans les entreprises et les réservoirs de documents numériques est issue des opérations de scannérisation ou des solutions GED/GEIDE. A part le fait de voir généralement écartée l'alternative de référencement par métadonnées normalisées dans les applications utilisées (i.e. Acrobat, Taurus), ces documents restent souvent exploités dans leurs versions graphiques. Une alternative de moindre gâchis est de procéder à une océarisation de ces ressources numériques pour les convertir en documents textuels. Ceci permettrait ensuite la possibilité de les indexer pour les intégrer dans les dépôts de documents numériques interrogeables.

La reconnaissance optique des caractères arabes n'est certes pas encore optimisée. Encore très peu de solutions arrivent à des taux élevés de reconnaissance. Moins sont encore ceux qui travaillent sur l'écriture arabe manuscrite.

### *Conclusion*

L'unification de la démarche éditoriale des documents numériques est une tache ardue voire impossible. Chaque institution et chaque structure dispose de ses propres besoins et de ses propres spécificités. L'objectif n'est donc pas de drainer les usages vers un stéréotype de modèle centralisé. Il s'agit plutôt d'une harmonisation de procédés, d'unification d'approches qui résulterait en une sorte de règles de « bonne conduite » capable de créer des fils conducteurs entre les pratiques observés parmi les producteurs des documents administratifs.

La convergence internationale vers les documents structurés en mode XML est sans doute la plus prometteuse vu sa neutralité envers les solutions logicielles et matérielles propriétaires. Mais devant l'ancrage des pratiques dans les modèles hérités des solutions propriétaires freinerait un certain temps la migration vers ces solutions universalistes.

Un grand chantier terminologique pour les effets d'indexation, de recherche par métadonnées normalisées reste encore à entamer. Mais l'entreprise est déjà rôdée à travers les systèmes ouverts et distribués, à travers les fonctions du workflow et du datamining qui sauront vite intégrer ces techniques et les rendre accessibles au grand public. Restera alors d'adapter tout le dispositif à des contraintes culturelles et linguistique parfois imposées par les usages socio-culturels voire même par des contraintes juridiques et procédurales.

# Notes